\documentclass[runningheads]{llncs}
\usepackage[T1]{fontenc}
\usepackage{graphicx,verbatim}

\usepackage{dsfont}
\usepackage{bm}

\usepackage{amsmath}
\usepackage{amssymb}

\usepackage{multirow}
\usepackage{tabularx}
\usepackage{booktabs}

\usepackage[table, xcdraw]{xcolor}
\usepackage{colortbl}
\usepackage{hyperref}
\usepackage{orcidlink}
\usepackage{glossaries}
\usepackage{fontawesome5}
\usepackage[ruled, vlined]{algorithm2e} 
\usepackage{color}
\usepackage{arydshln}
\begin{document}
%
\title{EchoLVFM: One-Step Video Generation via Latent Flow Matching for Echocardiogram Synthesis}
%

\author{Emmanuel Oladokun\inst{1}\orcidlink{0000-0003-2935-1552} \and   Sarina Thomas\inst{2}\orcidlink{0000-0002-1202-0856} \and  Jurica Šprem\inst{2}\orcidlink{0000-0002-9165-0847} \and  Vicente Grau\inst{1}\orcidlink{0000-0001-8139-3480}}  
\authorrunning{E. Oladokun et al.}
\institute{University of Oxford, Oxford, England \and GE HealthCare, Cardiovascular Ultrasound R\&D, Oslo, Norway\\ \email{emmanuel.oladokun@eng.ox.ac.uk}}
  
\maketitle              

\begin{abstract}
Echocardiography is widely used for assessing cardiac function, where clinically meaningful parameters such as left-ventricular ejection fraction (EF) play a central role in diagnosis and management. Generative models capable of synthesising realistic echocardiogram videos with explicit control over such parameters are valuable for data augmentation, counterfactual analysis, and specialist training. However, existing approaches typically rely on computationally expensive multi-step sampling and aggressive temporal normalisation, limiting efficiency and applicability to heterogeneous real-world data.

We introduce EchoLVFM, a one-step latent video flow-matching framework for controllable echocardiogram generation. Operating in the latent space, EchoLVFM synthesises temporally coherent videos in a single inference step, achieving a \textbf{$\sim$50×} improvement in sampling efficiency compared to multi-step flow baselines while maintaining visual fidelity. The model supports global conditioning on clinical variables, demonstrated through precise control of EF, and enables reconstruction and counterfactual generation from partially observed sequences. A masked conditioning strategy further removes fixed-length constraints, allowing shorter sequences to be retained rather than discarded.

We evaluate EchoLVFM on the CAMUS dataset under challenging single-frame conditioning. Quantitative and qualitative results demonstrate competitive video quality, strong EF adherence, and 57.9\% discrimination accuracy by expert clinicians which is close to chance. These findings indicate that efficient, one-step flow matching can enable practical, controllable echocardiogram video synthesis without sacrificing fidelity. Code available at:\href{https://github.com/EngEmmanuel/EchoLVFM}{\texttt{ \faGithub\,EchoLVFM}}
\keywords{Video Generation \and Ultrasound \and Echocardiography.}

\end{abstract}
\section{Introduction}
Echocardiography is a widely used cardiac imaging modality that supports diagnosis, monitoring, and treatment planning across a broad range of cardiovascular diseases \cite{Aly2021CardiacReview,Potter2019TheCare}. From echocardiogram videos, clinically meaningful physiological parameters such as left-ventricular ejection fraction (EF) can be derived, which play a central role in assessing cardiac function and diagnosing conditions including heart failure \cite{McDonagh20232023Failure}. As a result, echocardiography remains a cornerstone of routine clinical practice due to its non-invasive nature, low cost, and portability.

The ability to generate realistic echocardiogram videos while explicitly controlling such clinical parameters is highly desirable. Controllable synthesis enables the creation of counterfactual examples, allowing clinicians to visualise how changes in physiological variables may manifest in imaging. It also supports specialist training by exposing trainees to rarely observed pathological cases, and facilitates dataset rebalancing in settings where real-world data is skewed.

Generative modelling for echocardiography is, however, challenging. Compared to natural video data, public medical datasets are relatively small and heterogeneous, with substantial variation in sequence length, frame rate, and image quality. Early generative approaches based on variational autoencoders (VAEs) \cite{Kingma2013Auto-EncodingBayes} and generative adversarial networks (GANs) \cite{Goodfellow2014GenerativeNetworks} demonstrated the feasibility of medical image synthesis \cite{FriedrichDeepSynthesis,Oladokun2024TransesophagealModels}, but often suffered from over-smoothing, training instability, or limited diversity. More recently, diffusion models \cite{Ho2020DenoisingModels} have become the dominant paradigm for high-fidelity image \cite{Oladokun2025FromDiffusion} and video generation \cite{Ho2022VideoModels}, including latent video synthesis
\cite{Zhou2024HeartBeat:Models}, but their reliance on many iterative denoising steps leads to slow and computationally costly inference.

Flow matching \cite{Lipman2022FlowModeling} has recently emerged as a compelling alternative. By learning a deterministic transport between noise and data distributions, flow matching combines the stability and generative quality of diffusion models with the efficiency of normalising flows \cite{Kobyzev2021NormalizingMethods}. Recent work has shown that this framework can be further simplified to enable generation in as little as a single inference step \cite{Geng2025MeanModeling}, making it appealing for video generation where efficiency is critical.

Despite these advantages, applications of flow matching to echocardiography remain limited. Yazdani et al. \cite{Yazdani2025FlowQuality} apply linear flow matching for echocardiogram synthesis but restrict generation to key frames, leaving temporal dynamics unmodelled. Reynaud et al. \cite{Reynaud2025EchoFlow:Generation} extend flow matching to latent video generation conditioned on EF; however, their approach is limited to frame animation, relies on temporally normalised inputs obtained via cropping and resampling to a fixed number of frames, and requires $\sim$ 100 inference steps.

In practice, echocardiographic data is highly heterogeneous. Sequences vary substantially in length, frame rate, and quality depending on acquisition conditions. For example, the CAMUS dataset contains videos depicting only end-diastole to end-systole in as few as ten frames. Existing approaches \cite{Reynaud2024Feature-ConditionedSynthesis,Reynaud2025EchoFlow:Generation} simplify modelling by enforcing fixed sequence lengths, which either necessitates discarding shorter sequences or upsampling them via temporal interpolation. Both strategies risk losing information about the original temporal dynamics, limiting applicability to real-world clinical datasets. Our contributions are:
\begin{enumerate}
    \item We introduce \texttt{EchoLVFM}, the first one-step latent video flow-matching framework for echocardiogram generation, enabling temporally coherent video synthesis in a single inference step while preserving the sample quality.

    \item \texttt{EchoLVFM} enables controllable echocardiogram generation via global conditioning on clinically meaningful variables, demonstrated through precise control of left-ventricular EF, and supports reconstruction and counterfactual synthesis from partially observed sequences.

    \item We propose a masked video conditioning strategy that supports variable-length videos, eliminating the need for aggressive temporal normalisation and extending generation beyond single-frame animation to real-world clinical settings.
\end{enumerate}
\section{Methods}
\begin{figure}[tb]
    \centering
    \includegraphics[width=0.75\linewidth]{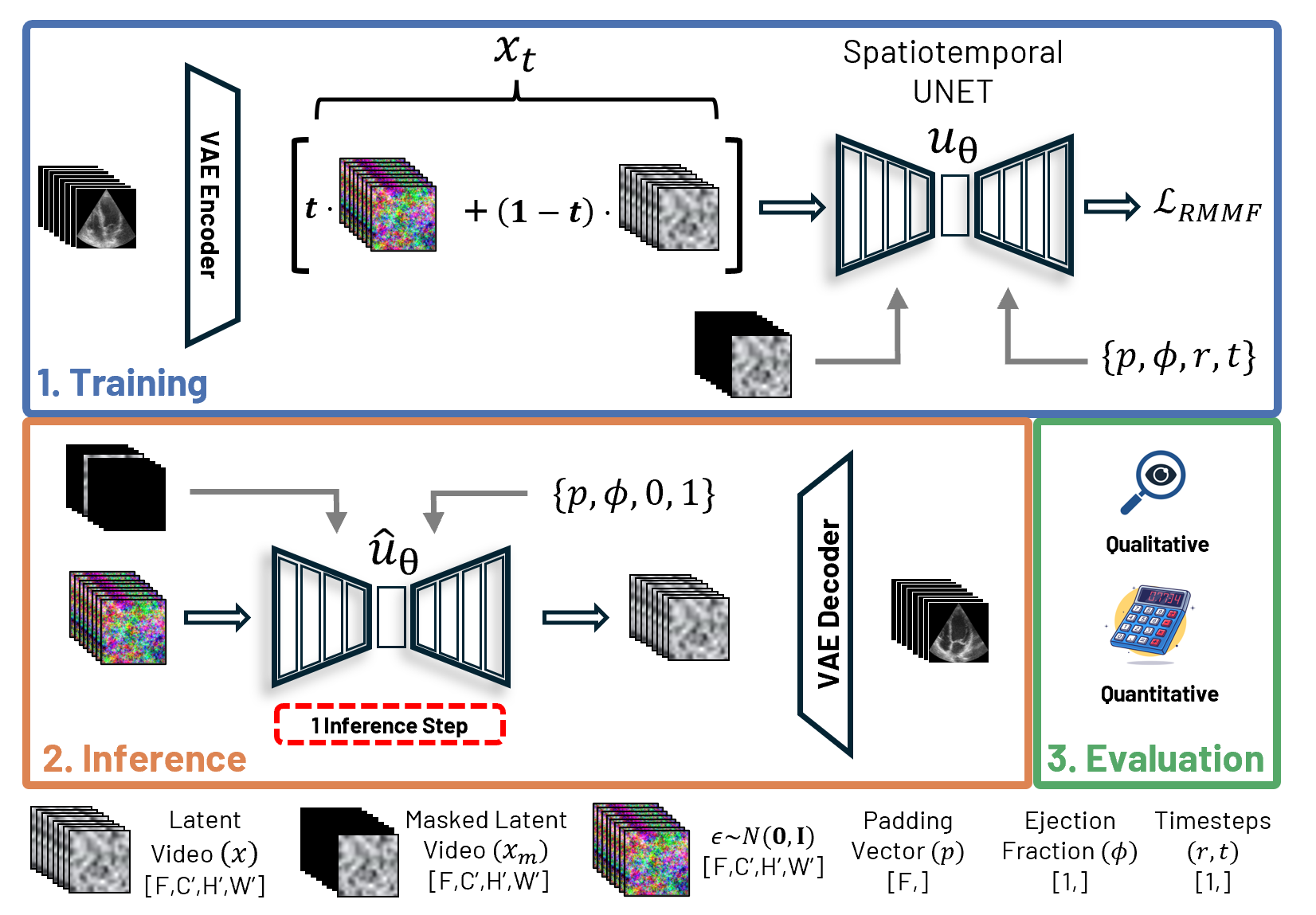}
    \caption{
    \textbf{EchoLVFM.} \textit{Training:}
    Videos are noised and processed in latent space. All but one observed frame are zeroed to form $x_m$, which serves as conditioning. To support variable-length sequences, a padding vector $p$ indicates which frames are valid observations in the temporally augmented input. The target EF $\phi$ is provided as global conditioning. $r$ and $t$ denote timesteps with $r<t$, and the model $u_{\theta}$ learns to predict the conditional average velocity over the interval $[r,t]$. \textit{Inference:}
    A partial video containing as little as a single observed frame, together with a target EF and random noise, is passed to the trained model. One-step integration produces the generated video.
    }
    \label{fig:fig1}
\end{figure}
\subsubsection{Flow Matching}
Flow matching learns a velocity field $v(x_t,t)=\frac{dx_t}{dt}$ that transports a data sample $x \sim p_{\mathrm{data}}$ to a noise sample $\epsilon \sim \mathcal{N}(0,I)$ as $t$ evolves from $0$ to $1$. Multiple interpolation paths can be defined between $\epsilon$ and $x$. In linear flow matching, the interpolation $x_t=(1-t)x+t\,\epsilon$ yields a constant target velocity $v(x_t,t)=\epsilon-x$, and a model $v_\theta$ is trained by minimising
$\mathbb{E}_{\epsilon,x,t}\|v_\theta(x_t,t)-(\epsilon-x)\|_2^2$.
At inference, the trajectory is obtained by solving the ODE, e.g.\ with Euler updates $x_{t+\Delta t}=x_t+\Delta t\,v_\theta(x_t,t)$. An alternative formulation, MeanFlow \cite{Geng2025MeanModeling}, models the \emph{average} velocity over an interval $[r,t]$,
$
u(x_t,r,t)=\frac{1}{t-r}\int_r^t v(x_\tau,\tau)\,d\tau,
$
which yields the MeanFlow identity
\begin{equation}
(t-r)u(x_t,r,t)=\int_r^t v(x_\tau,\tau)\,d\tau.
\label{eq:mean_flow_idt}
\end{equation}
Differentiating w.r.t.\ $t$ gives
$
u(x_t,r,t)
=
v(x_t,t)
-
(t-r)\frac{d}{dt}u(x_t,r,t).
$
Substituting $\frac{dx_t}{dt}=v(x_t,t)=\epsilon-x$ produces the effective regression target
\begin{equation}
u_{\mathrm{tgt}}\,
=\quad
v(x_t,t)
-
(t-r)\big(v(x_t,t)\partial_x u_\theta + \partial_t u_\theta\big) \quad=\quad (\epsilon-x) - \mathcal{I}(x_t, r, t).
\end{equation}
A network $u_\theta(x_t,r,t)$ can then be trained to regress $u_{\mathrm{tgt}}$ using the objective $\mathcal{L}_{\mathrm{MF}} = \mathbb{E}_{\epsilon,x,r, t} \|u_\theta(x_t,r,t) -\operatorname{sg} (u_{\mathrm{tgt}}) \|_2^2,$
where $\operatorname{sg}(\cdot)$ is the stop-gradient operator which prevents double backpropagation through the Jacobian–vector product. Once trained, sampling uses the mean velocity
$ x_r = x_t - (t-r)u(x_t,r,t),$ which can be used to perform one-step sampling like so: $x = \epsilon - u(\epsilon,0,1)$.
\subsubsection{EchoLVFM} 
We build upon MeanFlow by introducing \texttt{EchoLVFM} (Fig.~\ref{fig:fig1}), extending the framework to conditional latent video generation for controllable echocardiogram synthesis. EchoLVFM takes as input a partial video, potentially containing as little as a single observed frame, a target EF ($\phi$), and a padding indicator $p$, and generates a sequence of temporal length $f \leq F$, where $F$ denotes the maximum supported length.

Let $s \in \mathbb{R}^{f \times C \times H \times W}$ denote a sequence with $f$ frames. The sequence is encoded via a VAE into latent space and temporally normalised to length $F$, yielding $x \in \mathbb{R}^{F \times C' \times H' \times W'}$. If $f > F$, frames are uniformly downsampled; if $f < F$, zero-padding is applied along the temporal dimension, thus preserving shorter sequences. A masked latent counterpart $x_m$ is constructed by masking a subset of observed (non-padding) to be used for conditioning. A binary padding vector $p \in \{0,1\}^F$ indicates valid frames ($p_t=0$) and padded frames ($p_t=1$). The model predicts a conditional mean velocity $u_{\theta}(x_t,r,t;x_m,\phi,p)$ in the latent space. Henceforth, we denote the conditioning variables $\{x_m, \phi, p\}$ as $\mathbf{c}$.

Applying vanilla MeanFlow to videos of varying lengths introduces two issues. First, sequences with fewer valid frames contribute fewer terms to the loss, biasing optimisation towards longer videos. Second, padded frames contribute zero targets, encouraging the model to generate blank frames. Both effects intensify as $F$ increases, limiting the range of video lengths that can be generated.

To address this, we introduce a temporal loss mask $LM = 1 - p$ that excludes padded frames from optimisation. We incorporate $LM$ into the MeanFlow loss:
\begin{equation}
\mathcal{L}_{\mathrm{MMF}}
=
\mathbb{E}_{\epsilon,x,t,r}
\big[
\alpha\cdot\|M \odot e\|_2^2
\big],
\quad
\alpha = [(\textstyle\sum_{f=1}^{F} LM_f)C'H'W']^{-1},
\end{equation}
where $\hat u_{\theta}$ is the prediction, $e=\hat u_{\theta}-\operatorname{sg}(u_{\mathrm{tgt}})$ is the error, $M$ denotes the mask $LM$ broadcast to the shape of $e$, and $\alpha$ ensures that each video contributes equally to the loss function. Following \cite{Geng2024ConsistencyEasy}, we adaptively weight the loss function by a factor 
$w = \big(\| M \odot e \|_2^2 + \varepsilon\big)^{-h},$
with gradients stopped through $w$, yielding
\begin{equation}
\mathcal{L}_{\mathrm{MMF}}^{\mathrm{adapt}}
=
\mathbb{E}_{\epsilon,x,t,r}
[
\operatorname{sg}(w)\cdot
\alpha \cdot
\left\| M \odot e \right\|_2^2
].
\end{equation}
This weighting upscales gradient contributions when the residual is small, avoiding vanishing gradients as the model approaches the target and stabilising optimisation across noise levels. Recent analyses \cite{ZhangALPHAFLOW:MODELS,GengImprovedModels}  of the MeanFlow framework have highlighted training instability and optimisation challenges. To tackle this, we introduce a masked reconstruction regulariser. Using the predicted mean velocity $\hat{u}_{\theta}$, we calculate $\hat x = x_t - t \big(\hat u_{\theta}(x_t,r,t;\mathbf{c}) + \mathcal{I}(x_t,r,t) \big)$ and penalise
\begin{equation}
\mathcal{L}_{\mathrm{rec}}
=
\mathbb{E}_{\epsilon, x, t, r}
\left[
\alpha \cdot
\|M \odot (\hat x - x)\|_2^2
\right].
\end{equation}
The final \texttt{EchoLVFM} objective is the regularised masked mean flow loss
\begin{equation}
\boxed{
\mathcal{L}_{RMMF}
=
\mathcal{L}_{\mathrm{MMF}}^{\mathrm{adapt}}
+
\lambda_{\mathrm{rec}}\,
\mathcal{L}_{\mathrm{rec}}
}
\end{equation}
combining adaptive masked MeanFlow with reconstruction regularisation. 

Geng et al. \cite{Geng2025MeanModeling} reported that they achieved the best performance when alternating objectives during training, using linear flow matching $75\%$ of the time and Mean Flow $25\%$ of the time. To adapt this for our setting, we used a masked version of linear flow  matching $\mathcal{L}_{MLF}$. Concretely,
\begin{equation}
\label{eq:masked_fm_loss}
\mathcal{L}_{MLF}
=
\mathbb{E}_{\epsilon,x,t}
[
\alpha
 \cdot
\lVert
M \odot\, (\hat{v}_{\theta}(x_t, t; \mathbf{c})
-
(\epsilon - x))
\big\rVert_{2}^{2}
],
\end{equation}

Although the experiments of this work focus on the most challenging setting where only a single observed frame is present in $x_m$ at inference, EchoLVFM generalises beyond this regime. By interleaving an existing sequence with padded frames and reconstructing it, our method naturally performs temporal upsampling. Moreover, because the padding indicator $p$ is incorporated during training, the generated sequence length can be controlled up to $F$, enabling variable-length synthesis unlike prior approaches that always generate a video of fixed length.

\subsubsection{Ejection Fraction}
 For all videos, we assign a proxy EF using the area–length method \cite{Pujadas2004MRFunction}, computed directly from LV segmentation masks. Using the LV cavity area $A$ and long-axis length $L$, volume is approximated as $V=\frac{8}{3\pi}\frac{A^2}{L}$, yielding
$
EF \approx 1 - \frac{V_{\mathrm{ES}}}{V_{\mathrm{ED}}}
= 1 - \left(\frac{L_{\mathrm{ED}}}{L_{\mathrm{ES}}}\right)
\left(\frac{A_{\mathrm{ES}}}{A_{\mathrm{ED}}}\right)^2.
$
Each video is thus assigned a single proxy EF, used as the conditioning variable $\phi$ during training and inference. 
\section{Experiments}

\subsubsection{Data}

We use the CAMUS dataset \cite{Leclerc2019DeepEchocardiography}, comprising 1,000 echocardiogram videos from 500 patients, each providing apical two- and four-chamber views. We adopt the original patient-level split: 800 videos (400 patients) for training, 100 for validation, and 100 for testing. The sequences span ED to ES and are heterogeneous in both quality and temporal length (10–42 frames), reflecting realistic clinical acquisition variability. To enable latent video generation, publicly available echocardiogram VAEs were evaluated on a reconstruction task. The '4f4$_{[28,28,4]}$' model from \cite{Reynaud2025EchoFlow:Generation} achieved the best overall performance (FID = 5.16, FVD = 36.1, SSIM = 0.958, LPIPS = 0.0272, PSNR = 35.3, MAE = 1.00\%, RMSE = 1.73\%) and was therefore used in all experiments. All videos were resized to $112\times112$ and encoded into its latent space.

\subsubsection{Training}

We employ a conditional spatio-temporal UNet with four resolution levels and feature dimensions $\{128,128,256,256\}$, integrating self- and cross-attention to jointly model spatial structure and temporal dynamics under conditioning. The model contains 76.8M parameters. A maximum length $F=32$ was chosen. Following extensive hyperparameter exploration, the final configurations trained for 1000 epochs with $\lambda_{rec} = 1$, cosine annealing with an initial learning rate of $5\times10^{-5}$, and batch size 2. All experiments, including inference speed measurements, were conducted on a single NVIDIA L40s 48GB GPU. Flash attention with support for \texttt{torch.func.jvp} was not implemented at the time of our experiments. Consequently, we implemented a custom attention processor to retain memory-efficient attention, using the \texttt{jvp-flash-attention} \cite{Morehead2025JVPAttention} and \texttt{diffusers} \cite{vonPlaten2022Diffusers:Models} libraries.

\subsubsection{Evaluation}

We compared EchoLVFM against baselines trained solely with $\mathcal{L}_{MLF}$. Preliminary experiments showed Linear performance plateaued after 25 inference steps. Quantitative evaluation was conducted along three axes: \emph{efficiency}, \emph{video quality}, and \emph{EF adherence}, for both reconstruction (Rec) and generation (Gen). In Rec, the true EF was used for conditioning. In Gen, an EF differing by at least 5\% from the true value was sampled from the challenging range $[0,100]$, exceeding the training distribution and typical clinical values.

Sampling efficiency was measured and averaged over 100 iterations. Video quality was assessed using FID \cite{Heusel2017GANsEquilibrium}, FVD \cite{Unterthiner2018TowardsChallenges}, SSIM \cite{Wang2004ImageSimilarity}, and LPIPS \cite{Zhang2018TheMetric} at resolution $112\times112$. For robustness, three independent noise samples were generated per test video (300 samples total). To evaluate EF adherence, the LV cavity in generated videos was segmented using a pretrained nnUNet\cite{Isensee2021NnU-Net:Segmentation}, and the EF was computed. We report $R^2$, MAE, and RMSE between requested and observed EF. In line with prior work \cite{Reynaud2024Feature-ConditionedSynthesis}, we also report performance after rejection sampling, leveraging the independent EF estimator. Qualitative evaluation was performed via a blinded real-vs-fake assessment by two expert cardiologists (>15 years experience each). After calibration with real examples, they classified 120 videos (60 real, 60 generated from the best Linear and EchoLVFM models).
\section{Results \& Discussion}
\subsubsection{Quantitative Evaluation}

\begin{table}[tb]
\centering
\caption{\textbf{Quantitative Results.} Comparison of baseline and proposed methods. \texttt{Cond.} lists the inputs: Image (I) or single frame in $x_m$, Text (T), Motion Mask (MM), Video (V), and  partial video (pV). Classifier-free guidance (CFG). \texttt{Steps (Vid/$s$)} reports inference steps and video throughput. \texttt{Task} denotes reconstruction (Rec) and \colorbox[HTML]{E6EEF5}{generation (Gen)}. $pmf$ is the proportion of masked frames and $h$ is the exponent in adaptive weight $w$. $\mu$ denotes the mean score; (RS) denotes rejection sampling (three samples per conditioning, best retained). \textbf{Bold} and \textcolor{blue}{\textbf{Blue}} indicate best Rec and Gen performance. $\dagger$ Inference steps not reported (DDPM default 1000). $\ddag$ Clip length unspecified (likely FVD$_{16}$). 
}
\label{tab:table_1}
\resizebox{\textwidth}{!}{%
\begin{tabular}{@{}lcccccccclccc@{}}
\toprule
 &
   &
   &
   &
   &
  \multicolumn{4}{c}{Video Quality} &
   &
  \multicolumn{3}{c}{EF Adherence} \\ \cmidrule(lr){6-9} \cmidrule(l){11-13} 
\multirow{-2}{*}{Method} &
  \multirow{-2}{*}{Cond.} &
  \multirow{-2}{*}{\begin{tabular}[c]{@{}c@{}}Steps\\ \texttt{(Vid/$s$)}\end{tabular}} &
  \multirow{-2}{*}{Task} &
   &
  FID $\downarrow$ &
  FVD$_{10}$ $\downarrow$ &
  SSIM $\uparrow$ &
  LPIPS $\downarrow$ &
   &
  \begin{tabular}[c]{@{}c@{}}$R^2$, $\%$ $\uparrow$\\ $\mu$ (RS)\end{tabular} &
  \begin{tabular}[c]{@{}c@{}}MAE, $\%$ $\downarrow$\\ $\mu$ (RS)\end{tabular} &
  \begin{tabular}[c]{@{}c@{}}RMSE, $\%$ $\downarrow$\\ $\mu$ (RS)\end{tabular} \\ \midrule
HeartBeat\textsubscript{2C}\cite{Zhou2024HeartBeat:Models} &
  I, T &
  1000\textsuperscript{\dag} &
  - &
   &
  36.3 &
  9.6 \textsuperscript{\ddag} &
  0.61 &
  - &
   &
  \cellcolor[HTML]{EFEFEF}{\color[HTML]{EFEFEF} } &
  \cellcolor[HTML]{EFEFEF}{\color[HTML]{EFEFEF} } &
  \cellcolor[HTML]{EFEFEF}{\color[HTML]{EFEFEF} } \\
HeartBeat\textsubscript{4C}\cite{Zhou2024HeartBeat:Models} &
  I, T &
  1000\textsuperscript{\dag} &
  - &
   &
  36.0 &
  12.6 \textsuperscript{\ddag} &
  0.61 &
  - &
   &
  \cellcolor[HTML]{EFEFEF}{\color[HTML]{EFEFEF} } &
  \cellcolor[HTML]{EFEFEF}{\color[HTML]{EFEFEF} } &
  \cellcolor[HTML]{EFEFEF}{\color[HTML]{EFEFEF} } \\
CES\cite{Kondori2025ControlEchoSynth:Diffusion} &
  MM,V &
  1000 &
  - &
   &
   26.6 &
  69.6 \textsuperscript{\ddag} &
  0.57 &
  - &
   &
  \cellcolor[HTML]{EFEFEF}{\color[HTML]{EFEFEF} } &
  \cellcolor[HTML]{EFEFEF}{\color[HTML]{EFEFEF} } &
  \cellcolor[HTML]{EFEFEF}{\color[HTML]{EFEFEF} } \\ \midrule
 &
   &
   &
  Rec &
   &
  40.4 &
  153.9 &
  \textbf{0.73} &
  \textbf{0.128} &
   &
  \textbf{73 (84)} &
  \textbf{5.7 (4.0)} &
  \textbf{7.3 (5.6)} \\
\multirow{-2}{*}{Linear} &
  \multirow{-2}{*}{I, EF} &
  \multirow{-2}{*}{\begin{tabular}[c]{@{}c@{}}25\\ \texttt{(0.37)}\end{tabular}} &
  \cellcolor[HTML]{E6EEF5}Gen &
   &
  \cellcolor[HTML]{E6EEF5}42.2 &
  \cellcolor[HTML]{E6EEF5}154.2 &
  \cellcolor[HTML]{E6EEF5}- &
  \cellcolor[HTML]{E6EEF5}- &
   &
  \cellcolor[HTML]{E6EEF5}59 (67) &
  \cellcolor[HTML]{E6EEF5}\textcolor{blue}{\textbf{14.4}} (12.2) &
  \cellcolor[HTML]{E6EEF5}18.6 (16.8) \\ \hdashline
 &
   &
   &
  Rec &
   &
  46.8 &
  164.7 &
  \textbf{0.73} &
  0.129 &
   &
  49 (64) &
  7.8 (6.1) &
  10.1 (8.5) \\
\multirow{-2}{*}{Linear\textsubscript{CFG}} &
  \multirow{-2}{*}{I, EF} &
  \multirow{-2}{*}{\begin{tabular}[c]{@{}c@{}}25\\ \texttt{(0.37)}\end{tabular}} &
  \cellcolor[HTML]{E6EEF5}Gen &
   &
  \cellcolor[HTML]{E6EEF5}46.8 &
  \cellcolor[HTML]{E6EEF5}178.6 &
  \cellcolor[HTML]{E6EEF5}- &
  \cellcolor[HTML]{E6EEF5}- &
   &
  \cellcolor[HTML]{E6EEF5}6 (16) &
  \cellcolor[HTML]{E6EEF5}23.0 (21.3) &
  \cellcolor[HTML]{E6EEF5}28.2 (26.7) \\ \hdashline
 &
   &
   &
  Rec &
   &
  41.9 &
  169.9 &
  0.72 &
  0.133 &
   &
  42 (77) &
  8.1 (4.6) &
  10.7 (6.8) \\
\multirow{-2}{*}{Ours$^{\lambda_{rec} = 0}_{h=2}$} &
  \multirow{-2}{*}{I, EF} &
  \multirow{-2}{*}{\textbf{\begin{tabular}[c]{@{}c@{}}1\\ \texttt{(18.6)}\end{tabular}}} &
  \cellcolor[HTML]{E6EEF5}Gen &
   &
  \cellcolor[HTML]{E6EEF5}41.8 &
  \cellcolor[HTML]{E6EEF5}166.6 &
  \cellcolor[HTML]{E6EEF5}- &
  \cellcolor[HTML]{E6EEF5}- &
   &
  \cellcolor[HTML]{E6EEF5}53 (70) &
  \cellcolor[HTML]{E6EEF5}16.0 (11.7) &
  \cellcolor[HTML]{E6EEF5}20.0 (15.9) \\ \hdashline
 &
   &
   &
  Rec &
   &
  49.2 &
  162.1 &
  \textbf{0.73} &
  0.133 &
   &
  57 (80) &
  7.2 (4.2) &
  9.3 (6.3) \\ 
\multirow{-2}{*}{Ours $(h =1)$} &
  \multirow{-2}{*}{I, EF} &
  \multirow{-2}{*}{\begin{tabular}[c]{@{}c@{}}1\\  \texttt{(18.4)}\end{tabular}} &
  \cellcolor[HTML]{E6EEF5}Gen &
  \cellcolor[HTML]{E6EEF5} &
  \cellcolor[HTML]{E6EEF5}49.3 &
  \cellcolor[HTML]{E6EEF5}161.5 &
  \cellcolor[HTML]{E6EEF5}- &
  \cellcolor[HTML]{E6EEF5}- &
  \multicolumn{1}{c}{} &
  \cellcolor[HTML]{E6EEF5}\textbf{\textcolor{blue}{64}} (78) &
  \cellcolor[HTML]{E6EEF5}\textcolor{blue}{\textbf{14.4}} (10.8) &
  \cellcolor[HTML]{E6EEF5}\textcolor{blue}{\textcolor{blue}{\textbf{17.5}}} (13.7) \\ \hdashline
 &
   &
   &
  Rec &
   &
  \textbf{38.5} &
  \textbf{138.8} &
  0.70 &
  0.136 &
   &
  32 (75) &
  9.1 (4.9) &
  11.6 (7.0) \\ 
\multirow{-2}{*}{Ours $(h=2)$} &
  \multirow{-2}{*}{I, EF} &
  \multirow{-2}{*}{\begin{tabular}[c]{@{}c@{}}1\\ \texttt{(18.5)}\end{tabular}} &
  \cellcolor[HTML]{E6EEF5}Gen &
   &
  \cellcolor[HTML]{E6EEF5}\textbf{\textcolor{blue}{38.5}} &
  \cellcolor[HTML]{E6EEF5}\textbf{\textcolor{blue}{144.1}} &
  \cellcolor[HTML]{E6EEF5}- &
  \cellcolor[HTML]{E6EEF5}- &
  \multicolumn{1}{c}{\textbf{}} &
  \cellcolor[HTML]{E6EEF5}52 (74) &
  \cellcolor[HTML]{E6EEF5}16.3 (11.4) &
  \cellcolor[HTML]{E6EEF5}20.1 (14.9) \\ \midrule
 &
   &
   &
  Rec &
   &
  38.6 &
  155.2 &
  0.82 &
  0.098 &
   &
  93 (96) &
  3.1 (2.2) &
  3.8 (2.9) \\
\multirow{-2}{*}{\begin{tabular}[c]{@{}l@{}}Ours $(h=2)$\\ $pmf = 50\%$ \end{tabular}} &
  \multirow{-2}{*}{pV,EF} &
  \multirow{-2}{*}{\begin{tabular}[c]{@{}c@{}}1\\ \texttt{(18.5)}\end{tabular}} &
  \cellcolor[HTML]{E6EEF5}Gen &
   &
  \cellcolor[HTML]{E6EEF5}38.9 &
  \cellcolor[HTML]{E6EEF5}150.5 &
  \cellcolor[HTML]{E6EEF5}- &
  \cellcolor[HTML]{E6EEF5}- &
   &
  \cellcolor[HTML]{E6EEF5}-1 (7) &
  \cellcolor[HTML]{E6EEF5}25.5 (24.3) &
  \cellcolor[HTML]{E6EEF5}29.2 (28.1) \\ \bottomrule
\end{tabular}%
}
\end{table}
Unless stated otherwise, results correspond to the most challenging setting where only a single observed frame is present in $x_m$ at inference. The strongest diffusion-based baselines are included in Table~\ref{tab:table_1}; notably, these methods do not evaluate per-video EF adherence.

As shown in Table~\ref{tab:table_1}, EchoLVFM generates 18.5 videos per second using a single sampling step, compared to 0.37 videos per second for linear flow, which requires 25 steps, representing an approximate 50× improvement in efficiency. Despite this substantial reduction in inference cost, EchoLVFM achieves competitive and, in several cases, superior video quality metrics. The best-performing configuration, EchoLVFM$_{h=2}$, attains the lowest FID (38.5) for both reconstruction and generation, alongside the strongest FVD scores (138.8 Rec, 144.1 Gen). While the Linear model achieves slightly stronger perceptual similarity (LPIPS = 0.128), its distributional metrics remain higher (FID 40.4/42.2; FVD 153.9/154.2). These results demonstrate that a single-step EchoLVFM model can match or exceed the distributional quality of multi-step linear flow models while operating at a fraction of the computational cost, highlighting a favourable efficiency–quality trade-off. Ablating the reconstruction loss ($\lambda_{\mathrm{rec}}=0$) degrades performance, with FVD worsening by $\approx$ 30 points, highlighting its importance.

The nnUNet model used for LV segmentation achieved a Dice score of $93\%$ on the test set and was subsequently applied to generated videos for EF estimation. Table~\ref{tab:table_1} shows that Linear performs best in the reconstruction task, whereas EchoLVFM achieves the strongest performance in the generation task. Notably, generation is the more challenging setting: models are conditioned on a conflicting $x_m$ and evaluated on EF values extending beyond the training distribution. This suggests that while Linear excels at reconstructing inputs with their original EF, EchoLVFM$_{h=1}$ generalises more effectively under distributional shift.

EchoLVFM$_{h=2}^{pmf=50\%}$ shows when only 50\% of valid frames in $x_m$ are masked. In this setting, frame-level fidelity improves, as reflected by higher SSIM and lower LPIPS, while video-level dynamics degrade slightly, as indicated by increased FVD, with FID remaining stable. The strongest effects are observed in EF adherence, with $R^2=93\%$ in Rec but an expected drop to $-1$ in Gen, as $x_m$ acts as a confounder when conditioning on a new EF. This indicates that, while the previous results were obtained under the hardest setting, EF adherence in Rec improves substantially when additional frames are included in $x_m$.

Collectively, these findings demonstrate that EchoLVFM enables efficient, one-step, controllable video generation while maintaining competitive visual fidelity and strong clinical adherence under challenging conditioning regimes.

\begin{figure}[t]
    \centering
    \includegraphics[width=\linewidth]{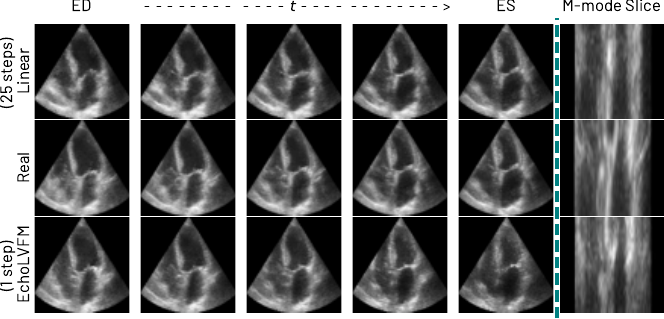}
    \caption{\textbf{Qualitative Results.} Columns 1-5 show frames sampled between and including ED and ES, while column 6 presents the M-mode slice of the middle row over time.}
    \label{fig:comparison}
\end{figure}

\subsubsection{Qualitative Results}
Across both clinicians, the overall confusion matrix was as follows: \{R as R: 59, R as S: 61, S as R: 40, S as S: 80\} where R and S correspond to Real and Synthetic, respectively. This corresponds to an overall accuracy of 58\%, compared with 50\% expected from random guessing in this binary task. Real videos were correctly identified 49\% of the time, indicating that clinicians struggled to reliably distinguish real echocardiograms from synthetic ones and suggesting strong perceptual realism in the generated videos. Comparing methods, synthetic videos produced using EchoLVFM$_{h=2}$ were identified as fake in 73\% of cases, and those produced using Linear were identified as fake in 60\% of cases. This suggests that Linear yields slightly more perceptually convincing outputs. However, EchoLVFM achieves generation in a single inference step, while Linear requires 25 steps, again illustrating the trade-off between efficiency and perceptual fidelity.
 
\section{Conclusion}
We introduced EchoLVFM, a one-step latent video flow-matching framework for controllable echocardiogram generation. By developing a novel loss, incorporating masked conditioning, and a padding indicator, our method removes the lower bound on usable sequence length, enabling shorter sequences to be retained rather than discarded. EchoLVFM supports conditioning on an arbitrary number of observed frames and naturally extends to tasks such as temporal upsampling.

Results show that EchoLVFM achieves competitive video quality and EF adherence in a single inference step, yielding substantial efficiency gains. While one-pass sampling has traditionally been a key advantage of GAN-based models, our results demonstrate that continuous-time generative models can now approach this level of efficiency without sacrificing stability or controllability. Our findings suggest that one-step flow matching provides a practical foundation for efficient, controllable modelling of realistic clinical video data.

\section{Acknowledgements}
The authors thanks Daria Kulikova and Anna Novikova for participating in the quiz.
We thank Phil Wang (lucidrains) \cite{2026PhilGitLab} for their implementation of the base Mean Flow method. The authors would like to acknowledge the use of the University of Oxford
Advanced Research Computing (ARC) facility (\url{http://dx.doi.org/10.5281/zenodo.22558})

%
%
%

\bibliographystyle{splncs04}
\bibliography{bibliography/miccai2026}
\end{document}